\documentclass[12pt]{article}
\usepackage{pic03}
\usepackage{hyperref}
\usepackage{url}
\usepackage{graphicx}
\usepackage{relsize}
\usepackage{xspace}
\usepackage{amssymb}
\def\babar{\mbox{\slshape B\kern-0.1em{\smaller A}\kern-0.1em
    B\kern-0.1em{\smaller A\kern-0.2em R}}}
\def\pep2{PEP-II}

\def\Km{K^-}
\def\Kp{K^+}
\def\piz{\pi^0}
\def\pip{\pi^+}

\def\Ds{D^+_s}
\def\DsTT{D_{sJ}^*(2317)^+}
\def\DsTO{D_s^{*}(2112)^+}
\def\Kbar{\kern 0.2em\overline{\kern -0.2em K}{}\xspace}
\def\DsFE{D_{sJ}(2458)^+}
\def\DmDg{\Delta m(\Ds\gamma)}
\def\DmDp{\Delta m(D_s^{*+}\piz)}

\newcommand{\gevc}{\ensuremath{{\mathrm{\,Ge\kern -0.1em V\!/}c}}\xspace}
\newcommand{\mevc}{\ensuremath{{\mathrm{\,Me\kern -0.1em V\!/}c}}\xspace}
\newcommand{\gevcc}{\ensuremath{{\mathrm{\,Ge\kern -0.1em V\!/}c^2}}\xspace}
\newcommand{\mevcc}{\ensuremath{{\mathrm{\,Me\kern -0.1em V\!/}c^2}}\xspace}

\begin{document}

\title{\bf Observation of New Narrow $\boldmath{D_s}$ states}

\author{
Antimo Palano       \\
{\em from the \babar\ Collaboration} \\ {\em INFN and University of Bari, Italy}}

\maketitle

%
%
%
%
%
%
\vspace{4.5cm}
%
\baselineskip=14.5pt
\begin{abstract}
The \babar\ experiment has discovered a new narrow state, $\DsTT$,  near 
2.32~\gevcc
in the inclusive $D_s^+ \piz$ invariant mass distribution
from $e^+e^-$ annihilation data at energies near 10.6~GeV~\cite{babar}. 
The same experiment has also shown evidence for structure in the 2.46~\gevcc
region in the $\DsTO \piz$ mass spectrum.
These discoveries 
have triggered several experiments in a search for new states coupled to 
the $D_s^+$ meson which confirmed the existence of $\DsTT$ together with
$\DsFE \to \DsTO \piz$ both in inclusive $e^+ e^-$ annihilation and in 
B decays. These two new
states are difficult to explain in terms of potential models.   
\end{abstract}
\newpage

\baselineskip=17pt

\section{Introduction}

Experimental information on the spectrum of the $c\overline{s}$ meson states
is limited. The $^1\!S_0$ ground state, the
$D_s^+$ meson, is well-established, as is the $^3\!S_1$ ground state,
the $\DsTO$. Only two other $c\overline{s}$ states are listed in the last
PDG edition~\cite{Hagiwara}. The $D_{s1}(2536)^+$ has been detected in
its $D^* K$ decay mode and analysis of the $D^*$ decay angular
distribution prefers $J^P = 1^+$.  
The $D_{sJ}^*(2573)^+$ was
discovered in its $D^0 \Kp$ decay mode and so has natural spin-parity.
The assignment $J^P=2^+$ is consistent with the data, but is not
established.  

The spectroscopy of $c\overline{s}$ states is simple in the limit of
large charm-quark mass~\cite{DeRujula,Isgur}.  
In that limit,
the total angular momentum $\vec{j}=\vec{l}+\vec{s}$ of the light
quark, obtained by summing its orbital and spin angular momenta,
is conserved.  The $P$-wave states, all of which have
positive parity, then have $j=3/2$ or $j=1/2$.  Combined with the
spin of the heavy quark, the former gives total angular momentum $J=2$
and $J=1$, while the latter gives $J=1$ and $J=0$.  
The $J^P=2^+$ and $J^P=1^+$ members of the $j=3/2$ doublet are
expected to have small width~\cite{Godfrey_91}, 
and are identified with the
$D_{sJ}^*(2573)^+$ and $D_{s1}(2536)^+$, respectively, although the latter 
may include a small admixture of the $j=1/2$, $J^P = 1^+$ state.
Theoretical models typically predict masses between
2.4 and 2.6~\gevcc for the remaining two
states~\cite{Godfrey_85,Godfrey_91,DiPierro}, both of
which should decay by kaon emission.
They would be expected to have large 
widths~\cite{Godfrey_91,DiPierro}
and hence should be difficult to detect. 

The experimental and theoretical status of the
$P$-wave $c\overline{s}$ states thus can be summarized by stating
that experiment has provided good candidates for the two states that
theory predicts should be readily observable, but has no candidates
for the two states that should be difficult to observe because
of their large predicted widths.

\section{The \babar\ Experiment.}
The \babar\ detector (at the \pep2
asymmetric-energy $e^+e^-$ storage ring with center-of-mass energy near 
10.6~GeV) is
a general purpose, solenoidal, magnetic spectrometer and is
described in detail elsewhere \cite{Aubert}. 
The data sample use in this analysis corresponds to an integrated luminosity
of 91~${\rm fb}^{-1}$.

\section{$D^+_s \pi^0$ events selection.}

The objective of this analysis is to investigate the inclusively-produced
$D_s^+ \piz$ mass spectrum by combining charged particles
corresponding to the decay $D_s^+\to \Kp \Km \pip$~\footnote{The 
inclusion of charge-conjugate 
configurations is implied throughout this paper.} with $\piz$ candidates
reconstructed from a pair of photons and performing a one-constraint fit to 
the $\piz$ mass.
A given event
may yield several acceptable $\piz$ candidates retaining only
those candidates for which neither 
photon belongs to another acceptable $\piz$ candidate.
To reduce combinatorial background from
the continuum and eliminate background from $B$-meson decay, each
$\Kp \Km \pip \piz$ candidate was required to have a center-of-mass momentum 
$p^*$ greater than 2.5~\gevc.

\begin{figure}[htb]
\begin{center}
\includegraphics[width=11cm]{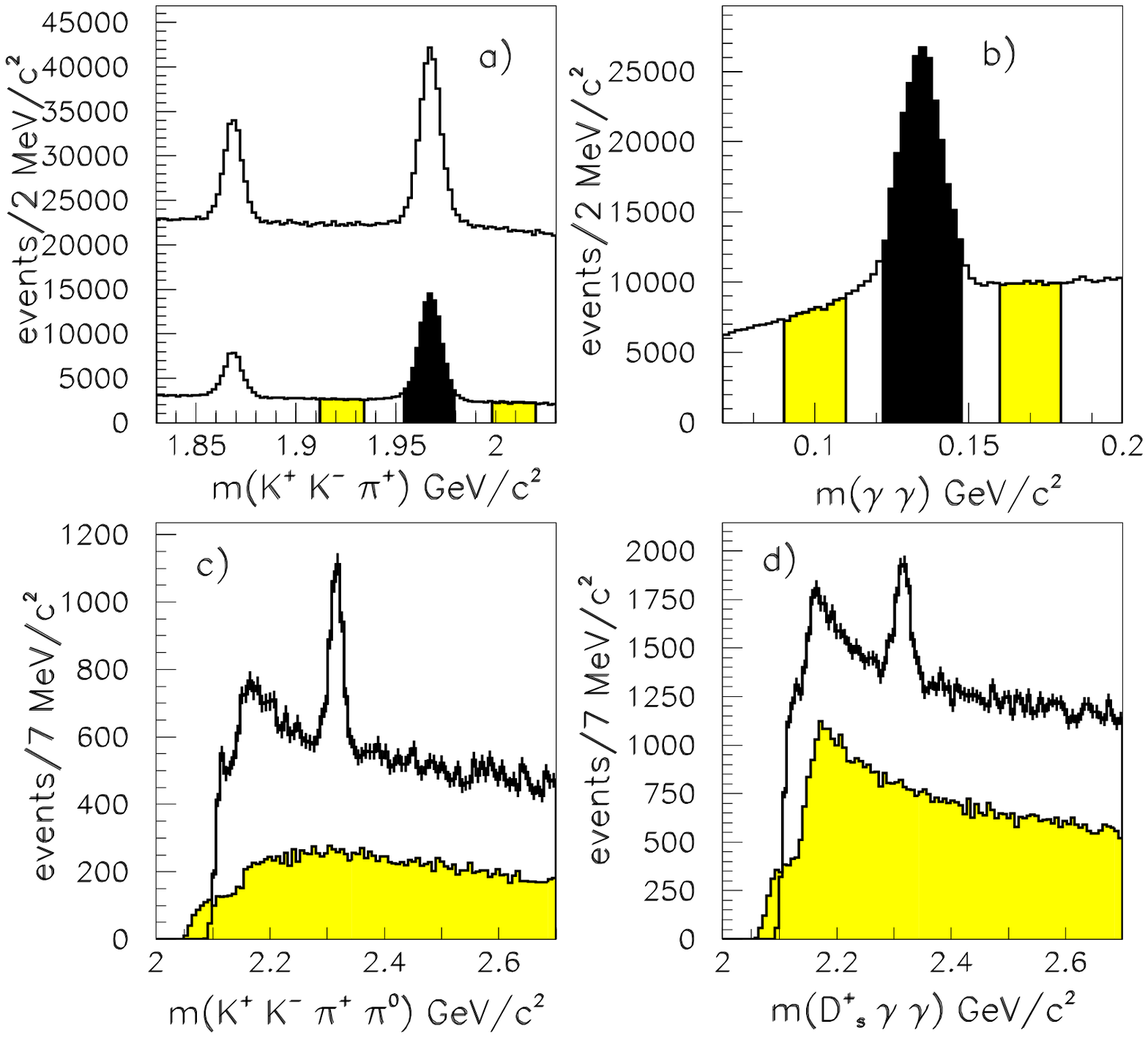}
\vspace{-0.5cm}
\caption{\label{fig:kkpi}
\babar\ experiment. 
(a) The distribution of $\Kp\Km\pip$ mass for all candidate events.
Additional selection criteria have been
used to produce the lower histogram.
(b) The two-photon mass distribution from $D_s^+\piz$ candidate events.
$D_s^+$ and $\piz$ signal and sideband regions are shaded.
(c) The $D_s^+\piz$ mass distribution for
candidates in the $D_s^+$ signal (top histogram) and $\Kp\Km\pip$
sideband regions (shaded histogram) of (a).
(d) The $D_s^+\gamma\gamma$ mass distribution for signal $D_s^+$
candidates and a photon pair from the $\piz$ signal region of (b) 
(top histogram) and
the sideband regions of (b) (shaded histogram). 
}
\end{center}
\end{figure}

The upper histogram in Fig.~\ref{fig:kkpi}(a) shows the $\Kp\Km\pip$ mass 
distribution for all candidates. Clear peaks corresponding to $D^+$ and $D_s^+$
mesons are seen. To reduce the background further,
only those candidates with $\Kp\Km$ mass
within 10~\mevcc of the $\phi(1020)$ mass
or with $\Km\pip$ mass within 50~\mevcc of the $\Kbar^*(892)$
mass are retained.
The decay products of the vector particles $\phi(1020)$ and 
$\Kbar^*(892)$ exhibit the
expected $\cos^2 \theta_h$ behavior required
by conservation of angular momentum, where $\theta_h$ is the
helicity angle. The signal-to-background ratio
is further improved by requiring $|\cos\theta_h|>0.5$. The lower histogram
of Fig.~\ref{fig:kkpi}(a) shows the net effect of these additional selection
criteria. The $D_s^+$ signal
and sideband regions are shaded.
The $D_s^+$ signal peak, consisting of approximately 80,000 events,
is centered at a mass of $(1967.20 \pm 0.03)$~\mevcc
(statistical error only).

Figure~\ref{fig:kkpi}(b) shows the mass distribution for all
two-photon combinations associated
with the selected events.  The $\piz$ signal 
and sideband regions are shaded.
Candidates in the $D_s^+$ signal region of Fig.~\ref{fig:kkpi}(a) are
combined with the 
mass-constrained $\piz$ candidates to yield the mass distribution
of Fig.~\ref{fig:kkpi}(c). A clear,
narrow signal at a mass near 2.32~\gevcc is seen.  The shaded histogram
represents the events in the $D_s^+\to\Kp\Km\pip$ mass sidebands combined with 
the $\piz$ candidates. 
In Fig.~\ref{fig:kkpi}(d)
the mass distributions result from the combination of the $D_s^+$
candidates with the photon pairs from the $\piz$ signal and 
sideband regions of Fig.~\ref{fig:kkpi}(b) (the sideband
distribution is again shaded). 
In Figs.~\ref{fig:kkpi}(c) and \ref{fig:kkpi}(d)
the 2.32~\gevcc signal is absent from the sideband distributions
indicating quite
clearly that the peak is associated with the $D_s^+\piz$ system.  
In order to improve mass resolution, the nominal
$D_s^+$ mass~\cite{Hagiwara} has been used to calculate the
$D_s^+$ energy.

\begin{figure}[htb]
\begin{center}
\includegraphics[width=8cm]{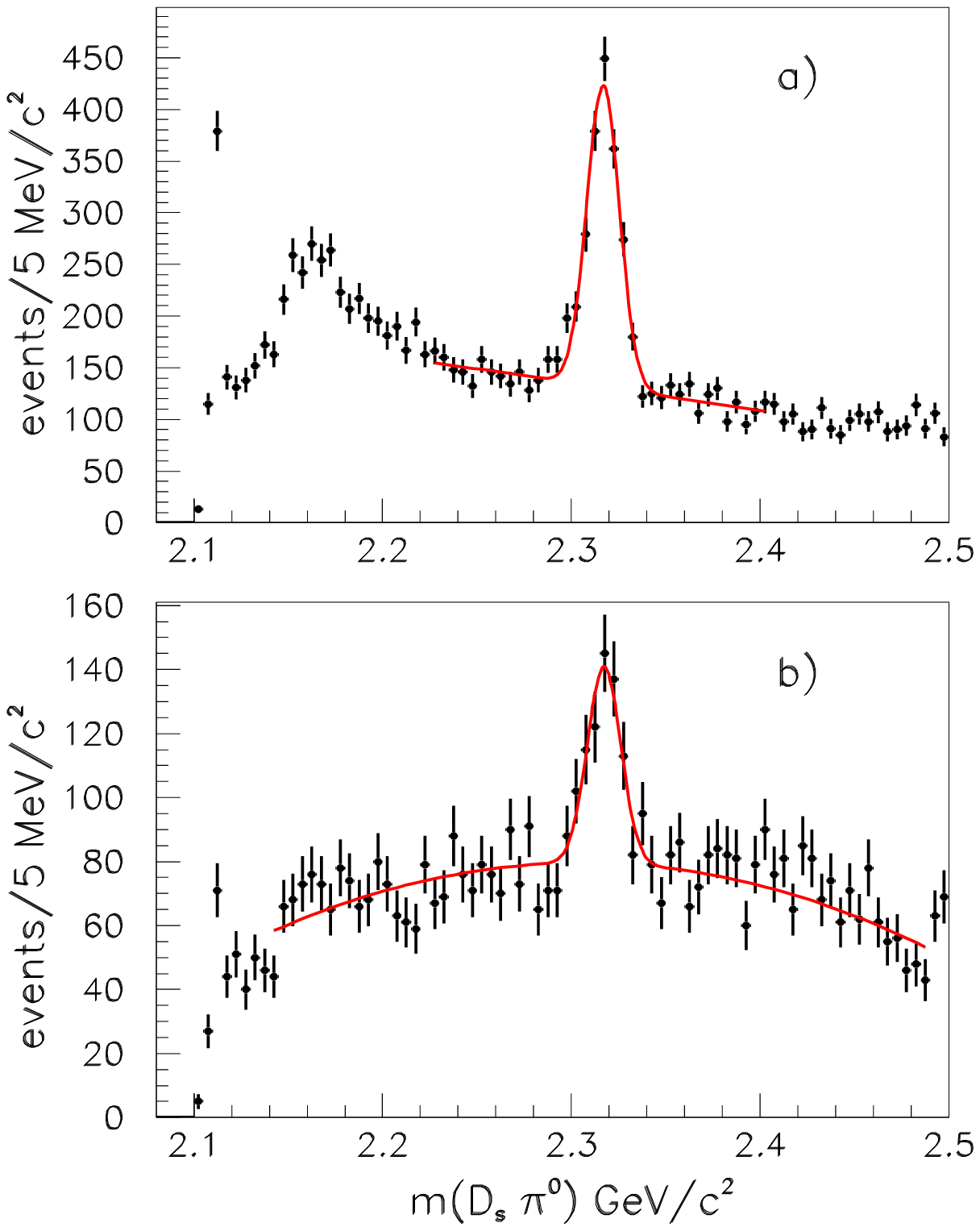}
\vspace{-0.5cm}
\caption{\label{fig:dspiz} 
\babar\ experiment. The $D_s^+\piz$ mass distribution
for (a) the decay $D_s^+  \to \Kp \Km \pip$
and (b) the decay $D_s^+  \to \Kp \Km \pip \piz$.
}
\end{center}
\end{figure}

The $D_s^+\piz$ mass distribution for $p^*(D_s^+\piz)>3.5$~\gevc
is shown in Fig.~\ref{fig:dspiz}(a). 
The fit function drawn on Fig.~\ref{fig:dspiz}(a) comprises a
Gaussian function describing the $2.32$~\gevcc signal and 
a polynomial background distribution function. 
The fit yields $1267 \pm 53$ candidates in the signal Gaussian with 
mass $(2316.8\pm 0.4)$~\mevcc and standard deviation
$(8.6\pm 0.4)$~\mevcc (statistical errors only).
The signal, labelled as $\DsTT$, is observed in both
the $\phi\pip$ and $\Kbar^{*0}\Kp$ decay modes of the $D_s^+$.
In addition, a sample of $D_s^+ \to \Kp\Km\pip\piz$ decays is
selected by adding $\piz$ candidates 
to each $\Kp\Km\pip$ candidate. 
Each resulting $D_s^+$ candidate is combined with a second $\piz$ candidate
with lab momentum greater than 300~\mevc. 
A clear $\DsTT$ signal is observed as shown in 
Fig.~\ref{fig:dspiz}(b). A Gaussian fit
yields $273 \pm 33$ events with a mean of $(2317.6\pm 1.3)$~\mevcc
and width $(8.8\pm 1.1)$~\mevcc (statistical errors only).
The mean and width are consistent with
the values obtained for the $D_s^+ \to \Kp\Km\pip$ decay mode. 
The mass distribution of the $D_s^+\to \Kp\Km\pip\piz$ sample (not shown) 
peaks at $(1967.4 \pm 0.2)$~\mevcc (statistical error only).

Monte Carlo simulations have been used to investigate the possibility that 
the $\DsTT$ signal could 
be due to reflection from other charmed states.
This simulation includes $e^+e^-\to c\bar c$ events 
and all known charm states and decays. The generated events were
processed by a detailed 
detector simulation and subjected to the same reconstruction
and event-selection procedure as that used for the data. No
peak is found in the 2.32~\gevcc $D_s^+\piz$ signal region.

Mass resolution estimates for the $\Kp\Km\pip\piz$ system are
obtained directly from the data using a fit to the mass distribution
$D_s^+\to \Kp\Km\pip\piz$. The measured width from this
mode is consistent
with that of the $\DsTT$ signal. It can be concluded that the
intrinsic width of the $\DsTT$ is small ($\Gamma \lesssim 10$~MeV).

\begin{figure}[htb]
\begin{center}
\includegraphics[width=8cm]{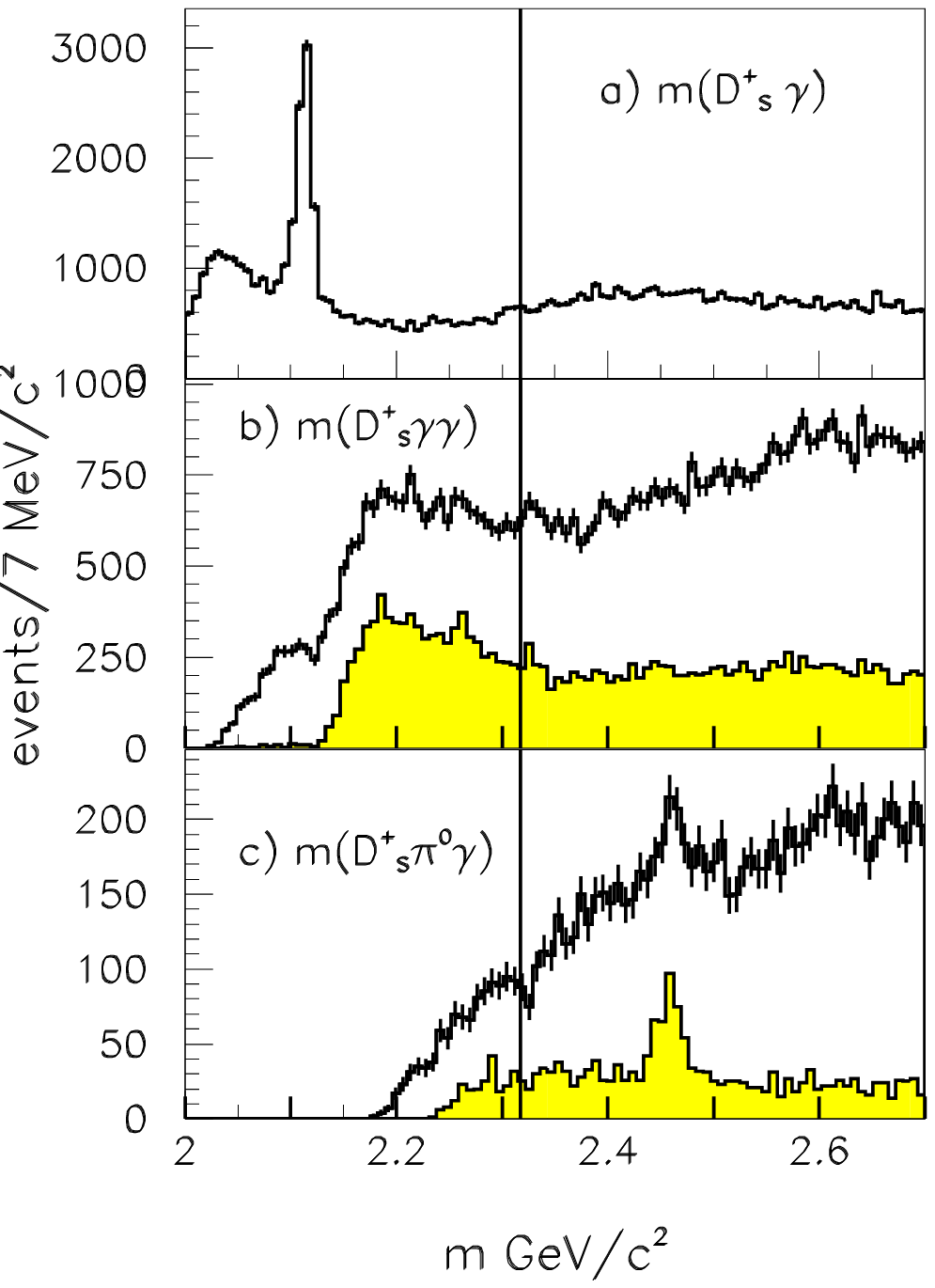}
\vspace{-3.0cm}
\caption{\label{fig:dsgammas}
\babar\ experiment. The mass distribution for (a)~$D_s^+\gamma$ and (b)~$D_s^+\gamma\gamma$
after excluding photons from the signal region of Fig.~\ref{fig:kkpi}(b).
(c)~The $D_s^+\piz\gamma$ mass distribution.
The lower histograms of (b) and (c) correspond to $D_s^+\gamma$
masses that fall in the $\DsTO$ signal region as described in
the text. The vertical line indicates the $\DsTT$ mass.}
\end{center}
\end{figure}

A search has also been performed for the decay 
$\DsTT \to D_s^+\gamma$.  
Shown in Fig.~\ref{fig:dsgammas}(a) is the $D_s^+\gamma$ mass 
distribution obtained by combining a $D_s^+$ candidate
in the signal region of Fig.~\ref{fig:kkpi}(a) with a photon
with an energy of at least 150~MeV
that does not belong to a $\gamma\gamma$ combination in the
signal region of Fig.~\ref{fig:kkpi}(b). The requirement
that the $p^*$ of the $D_s^+\gamma$ system be greater than
3.5~\gevc is also imposed. There is a clear $\DsTO$ signal,
but no indication of $\DsTT$ production.
The $D_s^+ \gamma\gamma$ mass distribution for 
$p^*(D_s^+ \gamma\gamma) > 3.5$~\gevc,
excluding any photon that belongs to the $\piz$ signal
region of Fig.~\ref{fig:kkpi}(b),
is shown as the upper histogram of 
Fig.~\ref{fig:dsgammas}(b). 
No signal is observed near 2.32~\gevcc. The shaded histogram
corresponds to the subset of combinations for which either $D_s^+ \gamma$
combination lies in the $\DsTO$ region, defined as 
$2.096 < m(D_s^+ \gamma) < 2.128$~\gevcc. Again, no $\DsTT$ signal 
is evident, thus demonstrating
the absence of a $\DsTO \gamma$ decay mode at the present level of
statistics. 
The $D_s^+ \piz \gamma$ mass distribution,
excluding any photon that belongs to any $\piz$ candidate,
is shown as the upper histogram of 
Fig.~\ref{fig:dsgammas}(c). 
The shaded histogram corresponds to the subset of combinations
in which the $D_s^+ \gamma$ mass falls in the $\DsTO$ region.
No signal is observed near 2.32~\gevcc in either case.
A small peak, however, is visible near a mass of 2.46~\gevcc. 
This mass corresponds to the
overlap region of the $\DsTO \to D_s^+ \gamma$ and 
$\DsTT \to D_s^+ \piz$ signal bands
that, because of the small widths of both the $\DsTO$
and $\DsTT$ mesons, produces a narrow peak in the
$D_s^+ \piz \gamma$ mass distribution that survives
a $\DsTO$ selection.

If the peak in the $D_s^+ \piz \gamma$ mass 
distribution of Fig.~\ref{fig:dsgammas}(c) were
due to the production of a narrow state with mass near 2.46~\gevcc
decaying to $\DsTO \piz$, the kinematics are such that a peak would
be produced in the  $D_s^+\piz$ mass distribution at a mass
near 2.32~\gevcc. Such a $D_s^+\piz$ mass peak, however, would
have a root-mean-square of $\sim 15$~\mevcc, which is significantly larger
than that obtained for the $\DsTT$ signal.
In addition, Monte Carlo studies indicate that if the apparent
signal at 2.46~\gevcc were due to a state that decays entirely
to $\DsTO\piz$, it would produce only one-sixth of the observed signal
at 2.32~\gevcc.

The \babar\ experiment was somewhat coutious in claiming the discovery of 
this second state.
``Although we rule out the decay of a state of mass 2.46~\gevcc
as the sole source of the $D_s^+\piz$ mass peak corresponding to
the $\DsTT$, such a state may be produced in addition
to the $\DsTT$. However, the complexity of the overlapping kinematics 
of the $\DsTO \to D_s^+\gamma$ and $\DsTT \to D_s^+ \piz$ 
decays requires more detailed study,
currently underway, in order to arrive at a definitive conclusion~\cite{babar}.''

\section{Confirmation of $\DsTT$ and observation of $\DsFE$ by other experiments.}

\begin{figure}[htbp]
  \centerline{\hbox{ \hspace{0.2cm}
    \includegraphics[width=11.0cm]{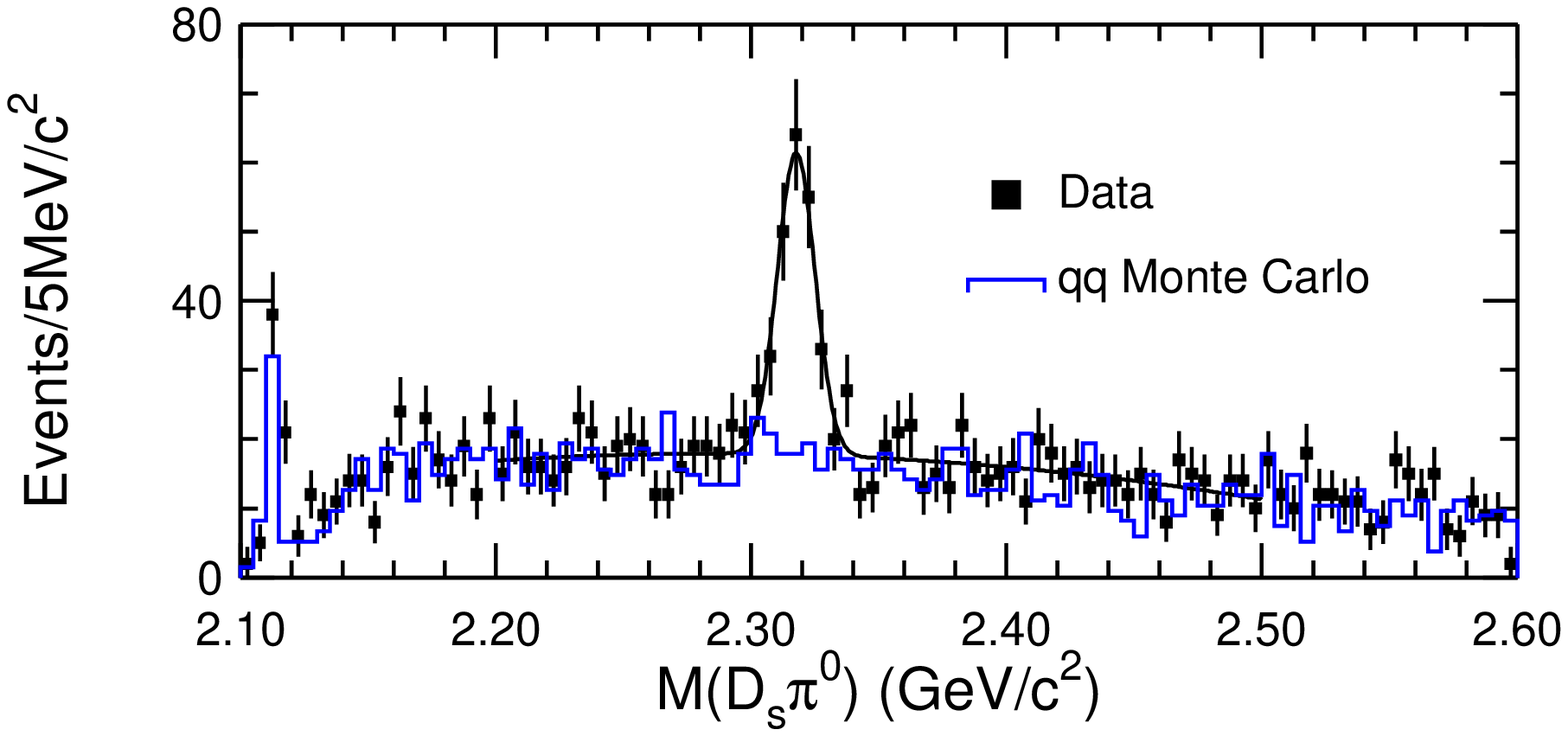}
    \hspace{0.2cm}
    \includegraphics[width=6.0cm]{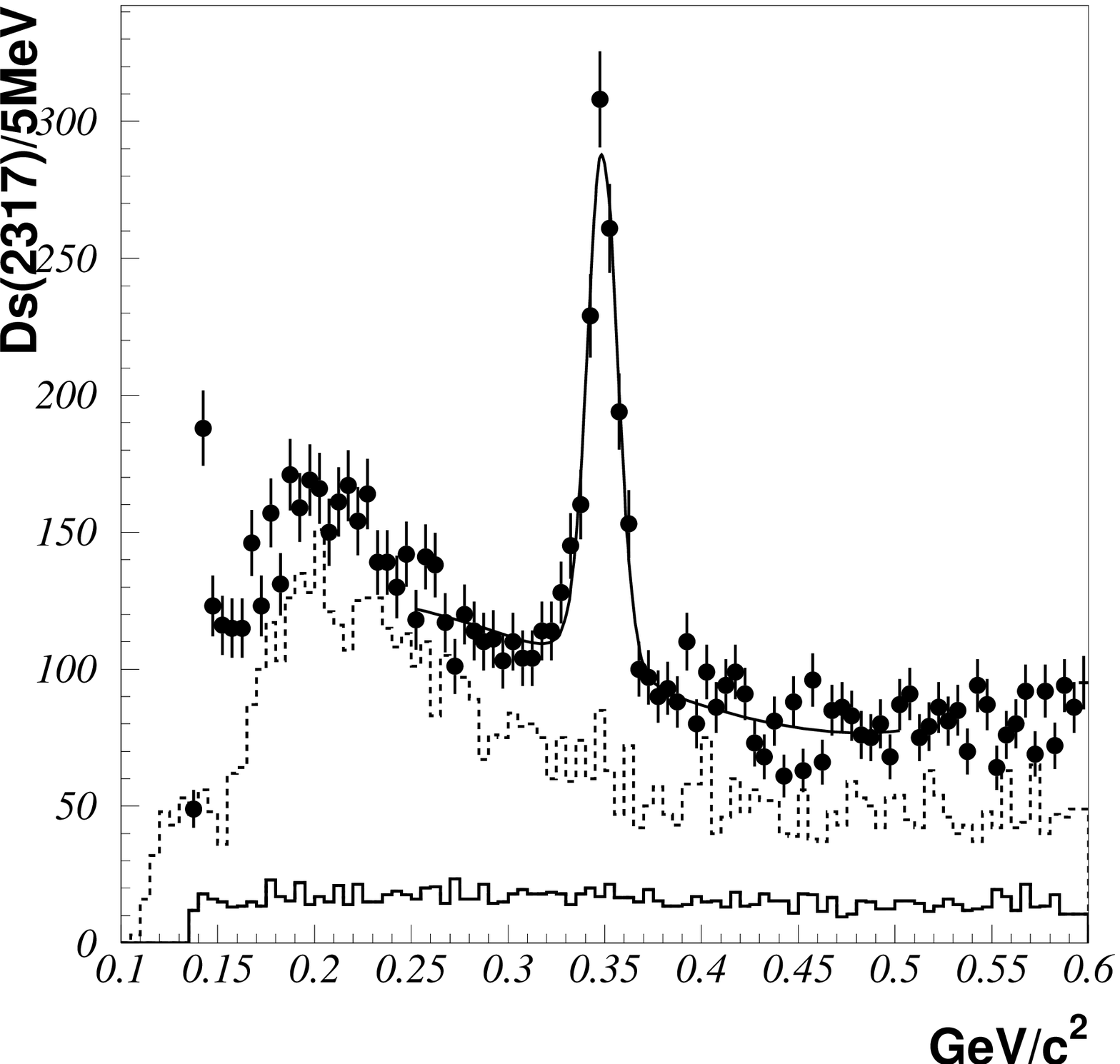}
    }
  }
\vspace{-0.5cm}
 \caption{\it
      $D^+_s \pi^0$ and $\Delta m$ from CLEO and Belle respectively showing 
the signal of $\DsTT$.
    \label{fig:cleo_belle} }
\end{figure}

Using 13.5 and 78 ${\rm fb}^{-1}$ integrated luminosity, the CLEO~\cite{cleo1} and 
Belle~\cite{belle1} experiments
 readily confirmed 
the existence of $\DsTT$ (shown in fig~\ref{fig:cleo_belle}). 
In terms of $\Delta m = m(K^+ K^- \pi^+ \pi^0) - m(K^+ K^- \pi^+)$, 
CLEO reported
a value of $\Delta m= 350.3 \pm 1.0$ \mevcc with 231 $\pm$ 30 events. 
Belle reports
$\Delta m= 348.9 \pm 0.5$ \mevcc with 643 $\pm$ 50 events. These values are in 
good agreement with the \babar\ measurement of $\Delta m=384.4 \pm 0.4$ \mevcc.

\begin{figure}[htbp]
  \centerline{\hbox{ \hspace{0.2cm}
    \includegraphics[width=7.0cm]{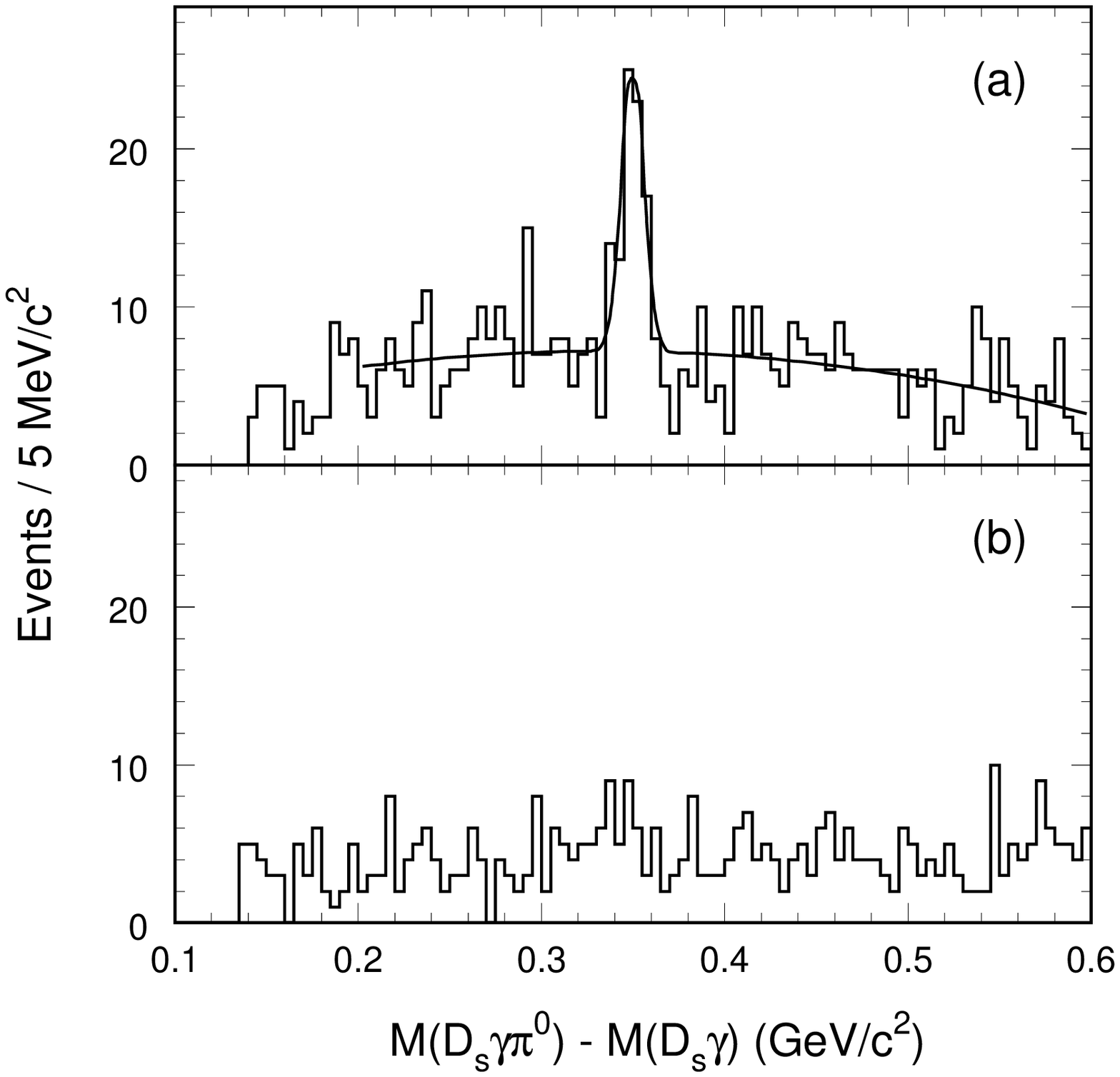}
    \hspace{0.2cm}
    \includegraphics[width=6.0cm]{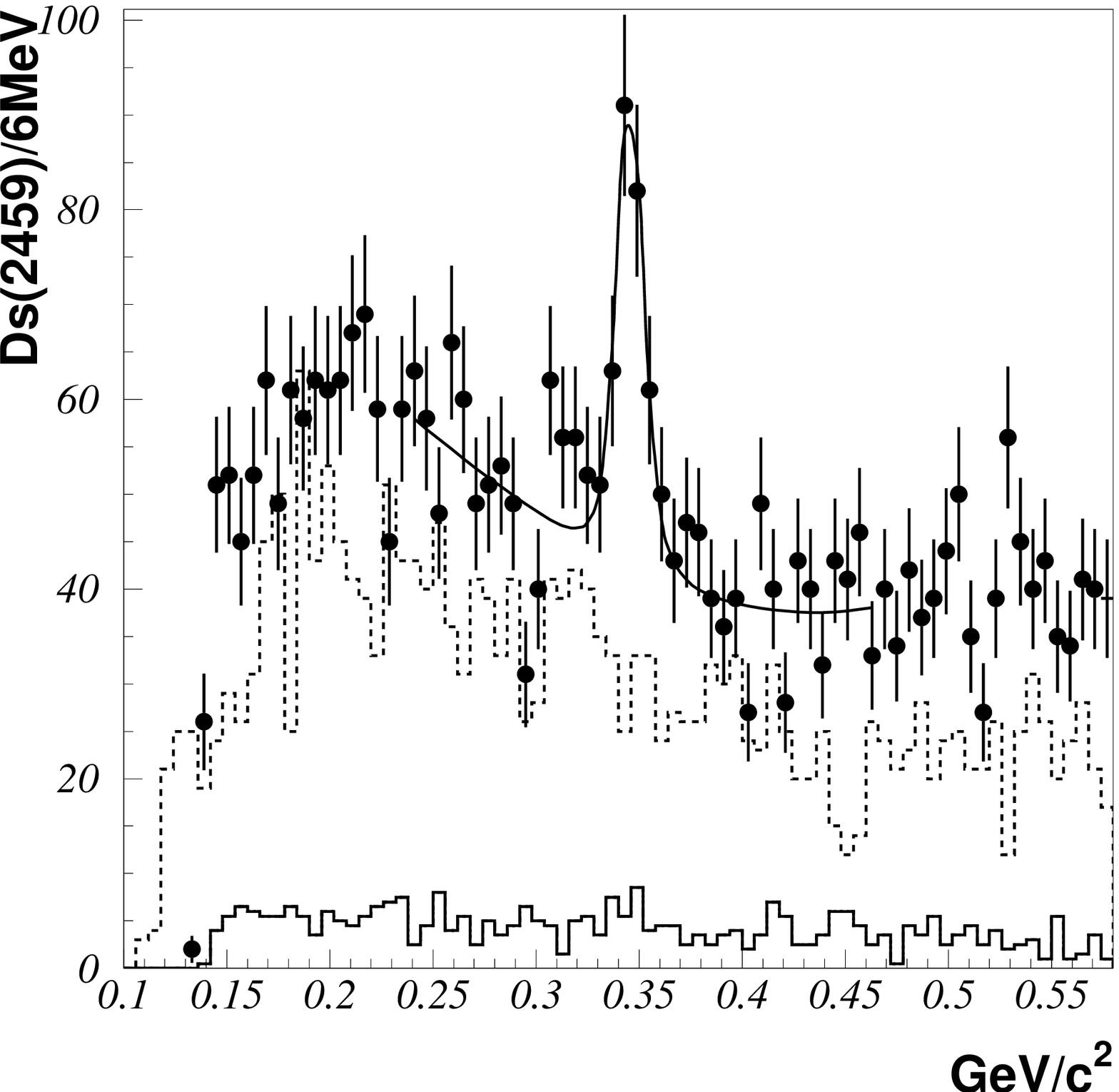}
    }
  }
\vspace{-0.5cm}
 \caption{\it
      $\Delta m (D_s^{*+} \pi^0)$ from CLEO and Belle Collaborations
showing the 
evidence for $\DsFE$. Data are from continuum $e^+ e^-$ annihilations.
The distribution from $D_s^{*+}$ sidebands is also shown.
    \label{fig:cleo_belle1} }
\end{figure}

In addition, both the CLEO and Belle Collaborations have analyzed the 
$\DsTO \piz$ mass 
distribution finding evidence for structure in the 2.46 GeV 
region (see fig.~\ref{fig:cleo_belle1}).
Defining now $\Delta m = m(K^+ K^- \pi^+ \pi^0 \gamma) - m(K^+ K^- \pi^+ \gamma)$, they reported the following parameters for this state:
$\Delta m$=351.2 $\pm$ 1.7 \mevcc with $41 \pm 12$ events (CLEO) and
$\Delta m$=344.1 $\pm$ 1.3 \mevcc with $79 \pm 18$ events (Belle).

Belle also reported evidence for both states~\cite{belle2} in the B decays
to $B \to D  \DsTT$ and  $B \to D  \DsFE$. In addition, they reported an 
observation of $\DsFE \to D^+_s \gamma$ in both B decays and continuum
$e^+ e^-$ annihilations. The angular analysis shows the expected behaviour for
a $J^P=1^+$ state.

No evidence has been found for narrow structures in $D^+_s \pi^\pm$ or 
$D^+_s \pi \pi$ final states.

\section{The observation of $\DsFE$ by the \babar\ experiment.}

In order to study the $D^+_s \pi^0 \gamma$ system, 
each $\Ds$ candidate is combined with all combinations of
accompanying $\piz$ candidates with momentum greater than 300~\mevc and
photon candidates of energy greater than 100~\mevcc. To suppress
background, photon candidates that belong to any $\piz$ candidate
are excluded and it is required that the combined
momentum $p^*$ in the $e^+e^-$ center-of-mass mass frame of each 
$\Ds\piz\gamma$ combination be greater than 3.5~\gevc.
\begin{figure}[htb]
\begin{center}
\includegraphics[width=11cm]{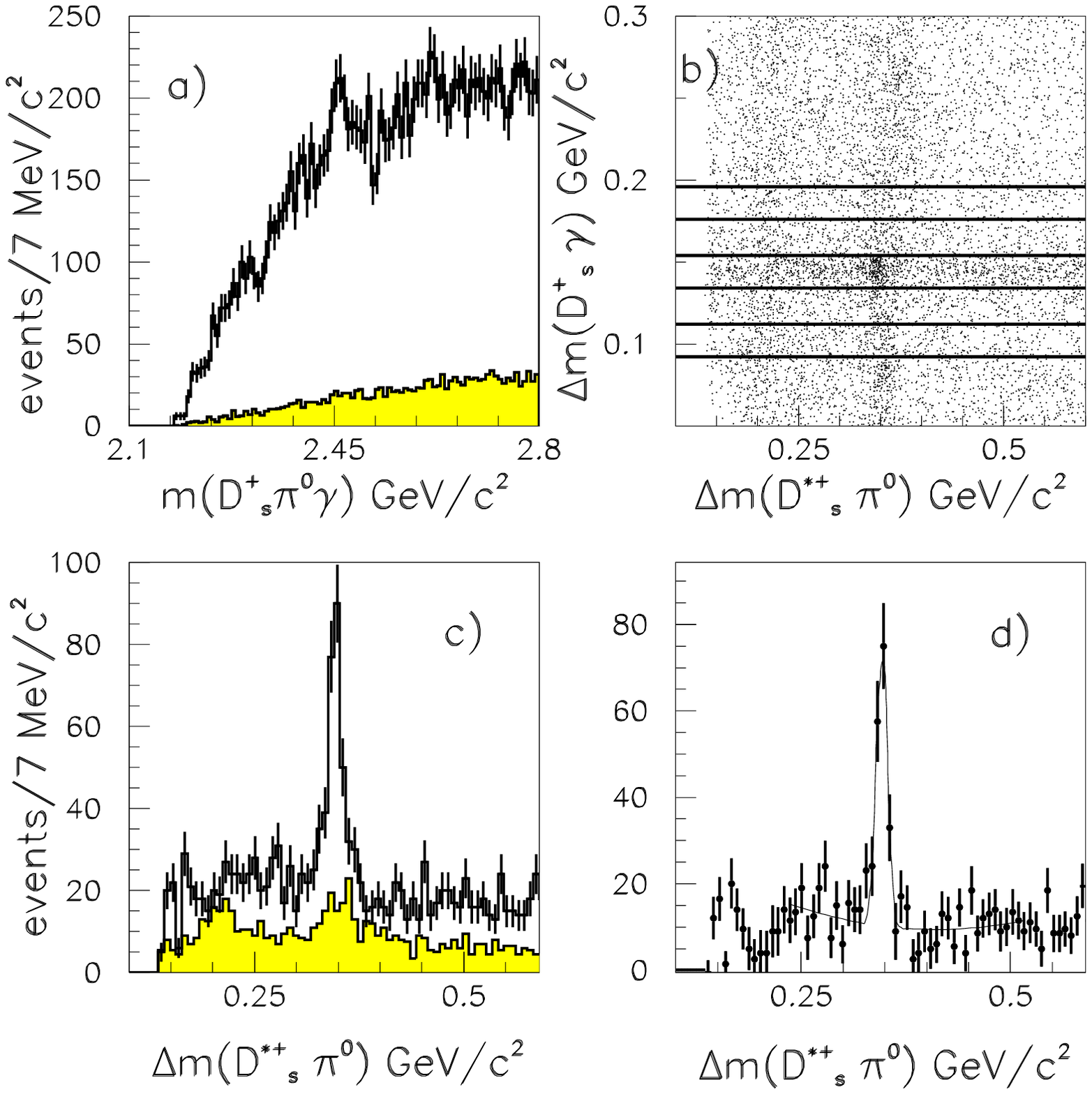}
\vspace{-0.5cm}
\caption{\label{fig:dspig}
\babar\ experiment. (a) The mass distribution for all selected $\Ds\piz\gamma$ 
combinations. The shaded region is from $\Ds$ sidebands defined by
$1.912<m(\Km\Kp\pip)<1.933$, $1.999<m(\Km\Kp\pip)<2.020~\gevcc$.
(b) The value of $\DmDg$ versus $\DmDp$  for all combinations.
The horizontal lines delineate three
ranges in $\DmDg$. 
(c) The $\DmDp$ mass distribution for the middle
range of $\DmDg$ (white) and for the upper and lower
ranges (shaded).
(d) The difference of the two histograms shown in (c).
The curve is the fit described in the text. 
}
\end{center}
\end{figure}
The $\Ds\piz\gamma$ invariant mass distribution is shown in 
Fig.~\ref{fig:dspig}. A small peak is observed near 2.46~\gevcc.
The background underneath this peak is from several sources, which
can be described in terms of mass differences defined as:
\begin{eqnarray}
\DmDg &\equiv& m(\Km\Kp\pip\gamma) - m(\Km\Kp\pip) \\
\DmDp &\equiv& m(\Km\Kp\pip\piz\gamma) - m(\Km\Kp\pip\gamma) \;.
\end{eqnarray}
A scatter plot of these quantities is plotted in Fig.~\ref{fig:dspig}b.
Two signals are clearly visible:
$\DsTO\to \Ds\gamma$ decay combined with unassociated $\piz$ candidates,
which appears as a horizontal band,
and $\DsTT\to\Ds\piz$ decay combined with unassociated $\gamma$ candidates,
which appears as a band that is almost vertical.

An enhancement is evident in the vicinity of the overlap of these two bands
with a $\Ds\piz\gamma$ mass near 2.46~\gevcc.  The upper histogram of
Fig.~\ref{fig:dspig}c shows events in the $\DsTO$ signal region, and the
shaded histogram shows those in the two $\DsTO$ sidebands. One can 
conclude that 
a state with decay to $\Ds\piz\gamma$ is seen to exist over a background from
$\DsTT$ combined with a $\gamma$. This background is
peaked at a mass slightly higher than the signal.  In
Fig.~\ref{fig:dspig}d the subtracted plot shows the narrow signal fitted
to a Gaussian shape on a second order polynomial background at
$\DmDg=346.2\pm 0.9$~\mevcc
(statistical errors only).

The signal, which is labelled $\DsFE$, may decay either through $\Ds\piz\gamma$,
$\DsTO\piz$ or $\DsTT\gamma$.  To disentangle these decay
modes and extract the most information from the data,
a binless-maximum likelihood fit to the $\Ds\piz\gamma$
system has been developed. 
The fit determines
a $\DsFE$ mass of ($2458.0 \pm 1.0$)~\mevcc and a measured Gaussian width
equal to ($8.5 \pm 1.0$)~\mevcc. 
This mass value (yielding $\Delta m = 345.6\pm 1.0$~\mevcc) agrees with 
that obtained by BELLE
but differs from that obtained by CLEO.

\begin{figure}[htb]
\begin{center}
\includegraphics[width=11cm]{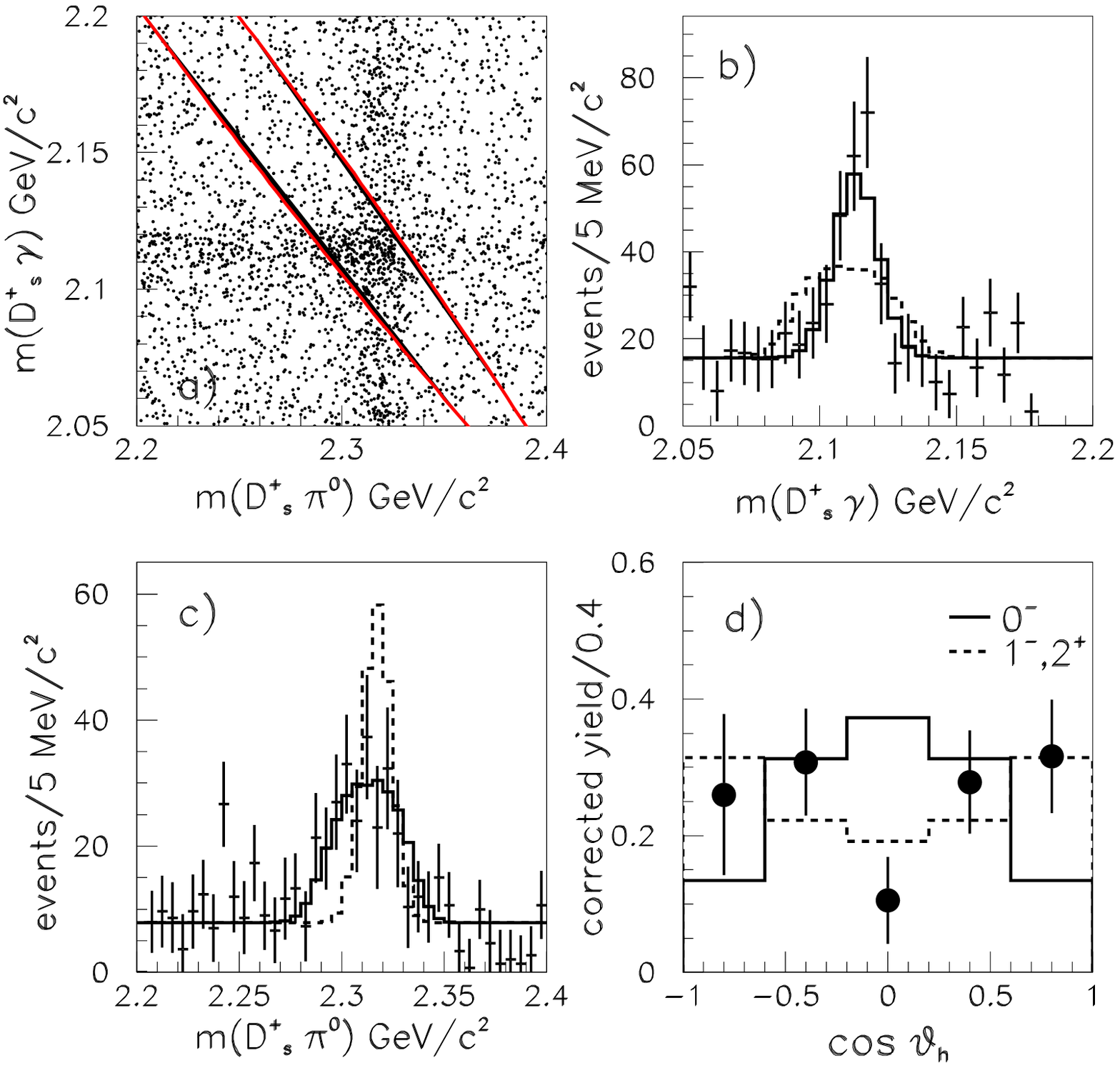}
\vspace{-0.5cm}
\caption{\label{fig:proj}
\babar\ experiment. (a) The $\Ds\piz$ versus $\Ds\gamma$ mass distribution for
all $\Ds\piz\gamma$ combinations. The curves indicate the
kinematically allowed region for $\DsFE$ decay.
(b) Sideband subtracted $\Ds\gamma$ mass distribution with
(line) Monte Carlo simulation for $\DsFE\to\DsTO\piz$ and
(dashed) $\DsFE\to\DsTT\gamma$. 
(c) A similar plot for the $\Ds\piz$ mass distribution.
(d) The efficiency corrected yield as a function of $\vartheta_h$
(statistical errors only).
The solid (dashed) histogram corresponds to the best fit of a
$\sin^2\vartheta_h$ ($1 + \cos^2\vartheta_h$) distribution.
}
\end{center}
\end{figure}

The shape of the $\Ds\piz$
and $\Ds\gamma$ distributions in the $\DsFE$ mass region 
can be used in order to
distinguish the two possible $\DsFE\to\DsTO\piz$ and 
$\DsFE\to\DsTT\gamma$ decay modes.
These shapes are influenced by the allowed kinematic range
for $\DsFE$ decay, as shown in Fig.~\ref{fig:proj}a.
Figs.~\ref{fig:proj}b--\ref{fig:proj}c show the $\DsFE$
background 
subtracted $\Ds\piz$ and $\Ds\gamma$ mass projections
compared with MC simulations of the two hypotheses.
The decay $\DsFE\to\DsTO\piz$ is clearly favored.

The distribution of the angle $\vartheta_h$ of the decay $\DsTO\to\Ds\gamma$ 
in its center-of-mass with respect to the $\DsFE$ can be used investigate the
spin-parity of the $\DsFE$. The resulting
efficiency-corrected $\cos\vartheta_h$ distribution is shown in
Fig.~\ref{fig:proj}d (statistical errors only). 
This distribution is not consistent with a
$\sin\vartheta_h$ distribution, which rules out a $\DsFE$ spin-parity
assignment of $J^P = 0^-$. 

MC simulations have been used to measure the detector resolution,
leading to the conclusion that the
intrinsic width of the $\DsFE$ is small ($\Gamma \lesssim 10$~\mevcc).

\section{Conclusions}

The decay of any $c\overline{s}$ state to $D_s^+\piz$ or $\DsTO \piz$ 
violates isospin conservation,
thus guaranteeing a small width for these states.
It is possible that the decays proceeds via $\eta-\piz$ mixing, 
as discussed by Cho and Wise~\cite{Cho}.

The low mass and the absence of a $\DsTT \to D_s^+ \gamma$
favors $J^P = 0^+$ for $\DsTT$. 
The mass of $\DsTT$ lies below the $DK$ threshold, the mass 
of $\DsFE$ lies above $DK$ and below $D^*K$ thresholds.

Different interpretations for these states have been proposed.
In ref.~\cite{Barnes, Beveren} models in terms of baryonia or molecules 
have been proposed. Ref.~\cite{Cahn} provides an explanation in terms 
of relativistic 
vector and scalar exchange forces. Ref.~\cite{Bardeen} uses HQET plus chiral
symmetry to predict parity doubling, i.e. the expectation is that the 
mass splitting between the $1^+$ and $1^-$ states should be the same as for
the $0^+$ and $0^-$ states.
    
Since a $c\bar{s}$ mesons with these masses contradicts
current models of charm meson 
spectroscopy~\cite{Godfrey_91,Godfrey_85,DiPierro}, 
most likely these models need modification.

\end{document}